\documentstyle[twocolumn,epsbox]{jpsj}

\def\runtitle{Strong Coupling Approach to the {\it d-p} Model
on the Basis of Fermi Liquid Theory}
\def\runauthor{Shigeru {\sc Koikegami} and Kosaku {\sc Yamada}}

\title
{
Strong Coupling Approach to the {\it d-p} Model \\
on the Basis of Fermi Liquid Theory \\
}

\author
{
Shigeru {\sc Koikegami}\footnote{E-mail: koike@ton.scphys.kyoto-u.ac.jp} 
and Kosaku {\sc Yamada}
}

\inst
{
Department of Physics, Kyoto University, Kyoto 606-8224\\
}

\recdate
{
December 10, 1997
}

\abst
{
We study the superconducting transition temperature \(T_{\rm c}\) of the
bilayer {\it d-p} model with \(d_{x^2-y^2}\)-wavelike attractive
interaction based on the formalism first employed by Nozi\`{e}res and
Schmitt-Rink. In the strong coupling regime, \(T_{\rm c}\) obtained 
through this formalism are much suppressed, compared with those through 
Thouless criterion only. We also find that, whether the interlayer 
coupling exsists or not, \(T_{\rm c}\) is almost propotional to 
the Fermi energy \(E_{\rm F}\) in the strong coupling regime. Thus, 
we can reproduce the essential nature in the underdoped region.
}

\kword
{high-\(T_c\) superconductor, bilayer {\it d-p} model, 
Nozi\`{e}res--Schmitt-Rink formalism, Thouless criterion}

\begin{document}
\sloppy
\maketitle

The pseudogap state of underdoped cuprates has been a very important issue 
in the study of high-\(T_{\rm c}\) superconductors (HTSC). 
Since the discovery of HTSC, the presence of an excitation gap, both 
in the charge and spin dynamics of underdoped cuprates in the normal 
state, has been indicated in several different experiments. 
The normal state transport properties (in-plane resistivity, Hall effect) 
and static susceptibility change their temperature dependences 
below a characteristic temperature \(T^*\).~\cite{rf:1} The electronic 
specific heat analysis has shown that the electronic entropy 
starts to decrease at the temperature well above \(T_{\rm c}\).~\cite{rf:2} 
The nuclear magnetic resonance (NMR) experiments have found the anomalous 
temperature dependences 
both of Knight shift and relaxation rate in almost all kinds of 
HTSC.~\cite{rf:3,rf:4} The suppression of the in-plane scattering rate well 
above \(T_{\rm c}\) has been derived from the optical conductivity 
mesurement.~\cite{rf:5} 

From all these various experimental results, we can deduce the existence 
of some kind of electronic bound states in the normal state of underdoped 
cuprates, whose characteristic energy should be related to \(T^*\). 
In order to explain this bound state formation, the theoretical works 
that identify \(T^*\) with the singlet resonating-valence-bond (RVB) 
formation temperture have been carried out earlier.~\cite{rf:6} In the 
recent years, angle resolved photo-emission spectroscopy (ARPES) 
has revealed the electronic structure of a pseudogap in the normal state, 
which has a highly anistropic momentum dependence, \(d_{x^2-y^2}\), 
as well as in the superconducting state.~\cite{rf:7,rf:8}

Based on this result, it may be natural that the origin of the pseudogap is 
in the preformed Cooper pair which cannot Bose-condensate due to 
its thermal fluctuation. The thermal fluctuation of the Cooper pair is 
neglected to determine \(T_{\rm c}\) in the original BCS theory, 
which is suitable for the description of the weak coupling regime 
where the electronic Fermi temperature \(T_{\rm F}\) is much larger 
than the temperature \(T\) and the superconducting carrier density is 
small compared to the degenerated electronic one. 
However, in underdoped cuprates, which are believed to be in the strong 
coupling regime where the electronic Fermi energy \(E_{\rm F}\) 
is much reduced by the strong correlation, the thermal fluctuation cannot 
be neglected over the wide temperature range above \(T_{\rm c}\), 
where \(T\) may exceed the Fermi temperature \(T_{\rm F}\). 

For the above reason, we use the formalism first employed by 
Nozi\`{e}res and Schmitt-Rink (NSR),~\cite{rf:9} in order to estimate 
\(T_{\rm c}\) of the bilayer {\it d-p} model in such a strong coupling 
regime. Their formalism and continuum models introduced in Ref. \citen{rf:9} 
have been adopted by the several authors in the past in order to study the 2- 
or 3-dimensional strong coupling superconductivity.~\cite{rf:10,rf:11,rf:12} 
The model Hamiltonian is as follows: 
\begin{eqnarray}
\lefteqn{H=H_++H_-+H_d}%
\nonumber\\ &   & \hspace{2em}%
- \mu \sum_{{\mib k} \sigma \tau}
(d_{{\mib k} \sigma \tau}^{\dagger}d_{{\mib k} \sigma \tau}
+p_{{\mib k} \sigma \tau}^{x \dagger}p_{{\mib k} \sigma \tau}^x
+p_{{\mib k} \sigma \tau}^{y \dagger}p_{{\mib k} \sigma \tau}^y).
\label{1}
\end{eqnarray}
Here \(d_{{\mib k} \sigma \tau}\)(\(d_{{\mib k} \sigma \tau}^{\dagger}\)) 
and \(p_{{\mib k} \sigma \tau}^{x(y)}\)
(\(p_{{\mib k} \sigma \tau}^{x(y)\dagger}\)) are the annihilation (creation) 
operator for \(d\)- and \(p^{x(y)}\)-electron of momentum \({\mib k}\), 
spin \(\sigma=\{\uparrow,\downarrow\}\) and layer \(\tau=\{+,-\}\), 
respectively. \(\mu\) is the chemical potential. 
The non-interacting part \(H_\pm\) is represented by
\begin{eqnarray}
H_\pm & = & \sum_{{\mib k} \sigma} \left( \begin{array}{ccc}
d_{{\mib k} \sigma \pm}^{\dagger} & p_{{\mib k} \sigma \pm}^{x \dagger}
& p_{{\mib k} \sigma \pm}^{y \dagger}\\
\end{array} \right)%
\nonumber \\ &   & \hspace{1em}%
\times%
\left( \begin{array}{lcc}
\varepsilon_d \mp t_z & \zeta_{\mib k}^x & \zeta_{\mib k}^y \\
-\zeta_{\mib k}^x & \varepsilon_p & \zeta_{\mib k}^p \\
-\zeta_{\mib k}^y & \zeta_{\mib k}^p & \varepsilon_p \\
\end{array} \right)
\left( \begin{array}{c}
d_{{\mib k} \sigma \pm} \\
p_{{\mib k} \sigma \pm}^x \\
p_{{\mib k} \sigma \pm}^y \\
\end{array} \right),
\end{eqnarray}
where \(\zeta_{\mib k}^{x(y)}=2{\rm i}t \sin \frac{k_{x(y)}}{2}\) and 
\(\zeta_{\mib k}^p=-4t_{pp} \sin \frac{k_x}{2} \sin \frac{k_y}{2}\). We 
introduce the three types of transfer energies, \(t\), \(t_{pp}\) and \(t_z\). 
The first two are the intralayer transfer energies for the nearest neighbors 
between \(d\)- and \(p^{x(y)}\)-orbitals and between \(p^{x}\)- and 
\(p^{y}\)-orbitals, respectively. The third is the interlayer transfer 
energy for between \(d\)-orbitals. In the following part of this letter, 
we take \(t\) as the unit of energy. The residual term \(H_d\) in 
eq. (\ref{1}) represents the attractive interaction between intralayer 
\(d\)-orbitals, described as follows: 
\begin{eqnarray}
H_d & = &  - \frac{U_d}{N} \sum_{{\mib k}{\mib k^\prime}}
g_{\mib k}g_{\mib k^\prime} %
\nonumber \\ &  & \hspace{4em}%
\times\sum_{{\mib q} \tau}%
d_{{\mib k} \uparrow \tau}^{\dagger}
d_{{\mib q}-{\mib k} \downarrow \tau}^{\dagger} 
d_{{\mib q}-\mib{k^\prime} \downarrow \tau}
d_{{\mib k^\prime} \uparrow \tau},
\end{eqnarray}
where \(g_{\mib k}=\cos k_x-\cos k_y\) is the \(d_{x^2-y^2}\)-wavelike form 
factor and \(N\) is the number of Cu sites. 

Then, we calculate the pairing correlation function for \(d\)-electrons. 
Diagonalizing \(H_++H_-\) by the unitary transformation,~\cite{rf:13} 
we can obtain the band dispersion, \(\varepsilon_{{\mib k}\pm}^{\alpha}\), 
and its weight for the \(d\)-electron, \(z_{{\mib k}\pm}^{\alpha}\), 
where \(\alpha=\{0,\pm1\}\) is the band index. They are defined by 
\begin{equation}
\varepsilon_{{\mib k}\pm}^{\alpha} = %
\frac{2}{\sqrt{3}}\,|t_{{\mib k}\pm}|\,\cos\,
\left(\frac{\varphi_{{\mib k}\pm}}{3}+\frac{2}{3}\pi\alpha\right) %
+\frac{\Delta_\pm}{3},
\end{equation}
with 
\begin{eqnarray}
t_{{\mib k}\pm}^2 & = & \frac{\Delta_\pm^2}{3}+(\zeta_{\mib k}^p)^2+%
|\zeta_{\mib k}^x|^2+|\zeta_{\mib k}^y|^2, \nonumber \\
\varphi_{{\mib k}\pm} & = & \frac{\pi}{2}+%
\left(\frac{\pi}{2}-\phi_{{\mib k}\pm}\right){\rm sgn}(s_{{\mib k}\pm}^3),%
\nonumber \\%
s_{{\mib k}\pm}^3 & = & \frac{\Delta_\pm^3}{27}%
+\Delta_\pm \left((\zeta_{\mib k}^p)^2-\frac{t_{{\mib k}\pm}^2}{3}\right)%
\nonumber \\ &   & \hspace{2em}%
-\zeta_{\mib k}^p(\zeta_{\mib k}^x\zeta_{\mib k}^{y*}%
+\zeta_{\mib k}^{x*}\zeta_{\mib k}^{y}),%
\nonumber \\%
\phi_{{\mib k}\pm} & = & \arctan\,\left(\left|s_{{\mib k}\pm}^6-\frac{4}{27}
t_{{\mib k}\pm}^6\right|^{1/2}/|s_{{\mib k}\pm}^3|\right),%
\nonumber \\%
\Delta_\pm & = & \varepsilon_d \mp t_z-\varepsilon_p,
\end{eqnarray}
and
\begin{eqnarray}
z_{{\mib k}\pm}^{\alpha} & = &
\frac{(\zeta_{\mib k}^p-\varepsilon_{{\mib k}\pm}^{\alpha})%
(\zeta_{\mib k}^p+\varepsilon_{{\mib k}\pm}^{\alpha})}%
{(\varepsilon_{{\mib k}\pm}^{\beta}-\varepsilon_{{\mib k}\pm}^{\alpha})%
(\varepsilon_{{\mib k}\pm}^{\alpha}-\varepsilon_{{\mib k}\pm}^{\gamma})},
\nonumber \\
\alpha & \neq & \beta, \hspace{1.5em} \beta \neq \gamma, \hspace{1.5em}
\gamma \neq \alpha.
\end{eqnarray}
Using these expressions, the intra-(inter-)layer pairing correlation 
function, \(\chi_{+(-)}^{dd}\), is represented by
\begin{eqnarray}
\lefteqn{\chi_\pm^{dd}({\mib q},\omega)}%
\nonumber\\  & = &
\frac{U_d}{2N_{\rm L}}\sum_{\mib k}g_{\mib k}^2\sum_{\alpha \beta \tau}%
\nonumber\\  &   & \hspace{2em}
\times%
\frac{z_{{\mib k}\tau}^{\alpha}z_{{\mib q}-{\mib k}\pm\tau}^{\beta}%
\{1-f(E_{{\mib k}\tau}^{\alpha})-f(E_{{\mib q}-{\mib k}\pm\tau}^{\beta})\}}%
{E_{{\mib k}\tau}^{\alpha}+E_{{\mib q}-{\mib k}\pm\tau}^{\beta}-\omega}.
\label{3}
\end{eqnarray}
Here \(E_{{\mib k}\pm}^{\alpha}= \varepsilon_{{\mib k}\pm}^{\alpha}-\mu\), 
and \(f(\omega)\) represents the conventional Fermi distribution function.
\(N_{\rm L}\) is the number of {\mib k}--space lattice points in the first 
Brillouin zone (FBZ). 
\begin{figure}
\vspace{10pt}
\epsfile{file=fig1,width=8cm,height=13cm}
\vspace{-10pt}
\caption{\(T_{\rm c}^{\rm MF}\) and \(T_{\rm c}^{\rm NSR}\) as functions 
of total electron number \(n\) in (a) for \(t_z=0.00t\) and in (b) for 
\(t_z=0.20t\). In both panels the closed symbols denote the results for 
\(T_{\rm c}^{\rm MF}\) and the open ones for \(T_{\rm c}^{\rm NSR}\). 
The circles, squares and diamonds correspond to \(\varepsilon_d=0.40t\), 
\(0.80t\) and \(1.20t\), respectively. The insets show the corresponding 
ratios of \(d\)-hole number \(n_{\rm h}^d\) to \(p\)-hole number 
\(n_{\rm h}^p\) for every \(\varepsilon_d\) as functions of \(n\). 
The correspondences of the symbols to \(\varepsilon_d\) are the same.}
\label{figure:1}
\end{figure}

NSR formalism takes into account the contribution of the
phase shift from the scattering \(t\)-matrix to the thermodynamic potential,
which leads to the equation that determines the critical temperature
\(T_{\rm c}^{\rm NSR}\) as the Bose-Einstein condensation
temperature for the density of pairs \(n-n_{\rm f}\), where \(n\) is
the total Fermion density and \(n_{\rm f}\) is the free part. First, the condition 
for the Cooper instability is given by 
\begin{eqnarray}
0 & = & 1-\chi_+^{dd}(0,0) \nonumber \\
  & = & 1-\frac{U_d}{2N_{\rm L}}\sum_{\mib k}g_{\mib k}^2\sum_{\alpha \beta \tau}%
\nonumber \\ &  & \hspace{3.5em}%
\times\frac{z_{{\mib k}\tau}^{\alpha}z_{{\mib k}\tau}^{\beta}%
\{1-f(E_{{\mib k}\tau}^{\alpha})-f(E_{{\mib k}\tau}^{\beta})\}}%
{E_{{\mib k}\tau}^{\alpha}+E_{{\mib k}\tau}^{\beta}},
\label{4}
\end{eqnarray}
and the equation which determines the chemical potential \(\mu\)
is described as
\begin{eqnarray}
n-n_{\rm f}\,(\mu,T_{\rm c}) & = & 2n_{\rm b}\,(\mu,T_{\rm c})%
\nonumber \\ & = & %
\frac{1}{N_{\rm L}}\sum_{{\mib q}\tau}g\,(\eta_{{\mib q}\tau}),%
\label{5}
\end{eqnarray}
where \(\eta_{{\mib q}\pm}\) is a discrete pole of
\([1-\chi_\pm^{dd}({\mib q},\omega)]^{-1}\) and \(g\,(\omega)\) represents 
the Bose distribution function. In order to subsequently define the Fermi 
energy, we introduce \(\mu^*\) as \(n_{\rm f}\,(\mu^*,T_{\rm c})=n\) for a 
given total density \(n\). Equation (\ref{4}) is called Thouless criterion, 
which results in the conventional BCS gap equation and gives the 
superconducting transition temperature evaluated in the mean-field 
thory \(T_{\rm c}^{\rm MF}\) if we neglect \(n_{\rm b}\) and substitute 
\(n=n_{\rm f}\) in eq. (\ref{5}). The summation of momentum and layer 
in eq. (\ref{5}) is executed, except for the point which 
satisfies eq. (\ref{4}), in order to remove the divergence. 
\begin{figure}
\vspace{10pt}
\begin{center}
\epsfile{file=fig2,width=8cm,height=7cm}
\end{center}
\vspace{-10pt}
\caption{\(U_d-\)dependence of \(T_{\rm c}^{\rm NSR}\) and 
\(T_{\rm c}^{\rm MF}\) for \(n=4.80\). The solid line and the dashed one 
denote \(T_{\rm c}^{\rm NSR}\) and \(T_{\rm c}^{\rm MF}\) respectively, and 
both are for \(t_z=0.00t\) and \(\varepsilon_d=1.20t\). 
We obtain these results by the interpolation based on the \(n\)-dependence 
of \(T_{\rm c}\) for every \(U_d\).}
\label{figure:2}
\end{figure}

\begin{figure}
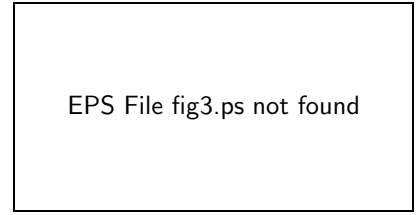
			
\vspace{12pt}
\begin{center}
\epsfile{file=fig3,width=8cm,height=7cm}
\end{center}
\vspace{-10pt}
\caption{\({\mib q}-\)dependence of \(\eta_{{\mib q}\pm}\) near \((0,0)\) 
for the two different fillings. These data have been reconstructed from the 
real numerical data by an appropriate method so that their general view can 
be easily observed. The solid line and the dashed one denote for 
\(n=4.1528\), \(T_{\rm c}^{\rm NSR}=0.01296t\) and for \(n=4.7198\), 
\(T_{\rm c}^{\rm NSR}=0.12625t\), respectively. Both are for \(t_z=0.00t\)
and \(\varepsilon_d=1.20t\). Arrows show the points for 
\({\mib q}=(q_{\rm c},0)\) and \({\mib q}=(0,q_{\rm c})\).}
\label{figure:3}
\end{figure}

\begin{figure}
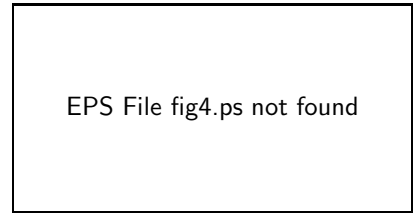

\vspace{10pt}
\epsfile{file=fig4,width=8cm,height=13cm}
\vspace{-10pt}
\caption{The ratios of \(T_{\rm c}(\equiv T_{\rm c}^{\rm NSR})\) to Fermi 
energy \(E_F\) in (a) for \(t_z=0.00t\) and in (b) for \(t_z=0.20t\) 
as functions of \(E_F/t\). 
In both panels the open circles, squares and diamonds correspond to 
\(\varepsilon_d=0.40t\), \(0.80t\) and \(1.20t\), respectively.}
\label{figure:4}
\end{figure}

In the numerical calculation, we set \(\varepsilon_p \equiv 0\) and control 
the filling by shifting \(\mu\). We define the Fermi energy 
\(E_{\rm F}\) as follows:
\begin{eqnarray}
E_{\rm F} \equiv \mu^* & - & E_{\rm F}^0, \nonumber \\
n_{\rm f}\,(E_{\rm F}^0,0) & = & 4.0,
\end{eqnarray}
which means that \(E_{\rm F}=0\) {\it i.e.} 
\(\mu^*=E_{\rm F}^0\) when the highest band is empty.
FBZ is divided into \(64 \times 64\) mesh, and the other input parameters 
are set like, \(\varepsilon_d=0.40t\), \(0.80t\), \(1.20t\), 
\(t_{pp}=0.40t\), \(U_d=4.0t\), \(t_z=0.00t\) for monolayer case or 
\(t_z=0.20t\) for bilayer case.
When we numerically integrate the sum in eq. (\ref{5}), we introduce
the cutoff momentum \(q_{\rm c}^\pm=(5/32)\pi\) assuming weak 
3-dimensionality of real systems, in order to avoid the large 
fluctuation which is unique to lower-dimensional system. 
In practice, first we investigate \(T_c^{\rm MF}\) for a given \(\mu\), 
then calculate \(n_{\rm b}\,(\mu,T_{\rm c})\) and determine \(\mu^*\) for the 
total density \(n\). Thus, the \(n_{\rm f}(\mu^*)\equiv n\) 
to \(T_{\rm c}^{\rm NSR}\) curve and the \(n_{\rm f}(\mu)\equiv n\) to 
\(T_{\rm c}^{\rm MF}\) curve can be obtained. Figure \ref{figure:1} shows 
these for the two differnt values of \(t_z\). The difference 
between \(T_{\rm c}^{\rm MF}\) and \(T_{\rm c}^{\rm NSR}\) increases much 
near the half-filled state, which corresponds to \(n=5.0\) for our model. 
While \(T_{\rm c}^{\rm MF}\) are higher for larger \(\varepsilon_d\), 
\(T_{\rm c}^{\rm NSR}\) take almost the same values near the half-filling, 
independent of \(\varepsilon_d\). It means that while \(T_{\rm c}^{\rm MF}\) 
are determined by \(U_d/W\), \(T_{\rm c}^{\rm NSR}\) are determined only by 
the filling \(n\), where \(W\) is the bandwidth of the highest band, 
which decreases as \(\varepsilon_d\) increases. In addition, the suppression 
of \(T_{\rm c}^{\rm NSR}\) is sensitive to the exsistence of the interlayer 
transfer \(t_z\), which is supposed to introduce the higher dimensionality 
and weaken the strong thermal fluctuation. In Fig. \ref{figure:2} we compare 
the \(U_d-\)dependence of \(T_{\rm c}^{\rm NSR}\) and \(T_{\rm c}^{\rm MF}\) 
for \(n=4.80\), \(t_z=0.00t\) and \(\varepsilon_d=1.20t\). 
While \(T_{\rm c}^{\rm MF}\) increases rapidly with \(U_d\), 
\(T_{\rm c}^{\rm NSR}\) is almost saturated in the strong coupling regime.
For references, in Fig. \ref{figure:3} we roughly show the momentum 
dependence of \(\eta_{{\mib q}\pm}\), which corresponds to the dispersion of 
a bosonic excitation {\it i.e.} a noncondensed Cooper pair, for two different 
fillings. \(\eta_{{\mib q}\pm}\) becomes lower as \(n\) approaches \(5.0\), 
and we expect that the densities of the noncondensed Cooper pairs are 
increased and that the Bose condensation temperatures {\it i.e.} 
\(T_{\rm c}^{\rm NSR}\) of the pairs are much suppressed, compared with 
the temperature where the Cooper instability occurs {\it i.e.} 
\(T_{\rm c}^{\rm MF}\). In Fig. \ref{figure:4}, we show the relation between 
\(T_{\rm c}^{\rm NSR}/E_{\rm F}\) and \(E_{\rm F}\) 
for each set of parameters. In our calculation, for every case 
\(T_{\rm c}^{\rm NSR}/E_{\rm F}\) shows a maximum in the 
intermediate coupling regime. This result is an artifact of approximations, 
and the method to remove such a maximum has been already 
discussed by R. Haussmann in his paper.~\cite{rf:14} However, in the strong 
coupling regime, all qualitative results are consistent independent of any 
kinds of approximations, and we find that \(T_{\rm c}^{\rm NSR}/E_{\rm F}\) 
tend to settle with constant values. 

In summary, we have calculated \(T_{\rm c}^{\rm NSR}\) of the bilayer 
{\it d-p} model on the assumption that \(d_{x^2-y^2}\)-wavelike attractive 
intraction works on \(d\)-electrons. 
Compared with \(T_{\rm c}^{\rm MF}\), \,\(T_{\rm c}^{\rm NSR}\)
are much suppressed in the strong coupling regime. This effect is due to 
the thermal fluctuation of preformed Cooper pairs and can be weakened 
by the exsistence of the interlayer coupling. Near the half-filled state, 
\(T_{\rm c}^{\rm NSR}\) is almost propotional to Fermi energy 
\(E_{\rm F}\). However, our model does not include the on-site repulsive 
interaction among \(d\)-electrons, which is very important to explain the 
Mott-insulating state near half-filling. In real systems, \(E_{\rm F}\) 
should be much reduced near half-filling by this strong correlation, where 
\(T_{\rm c}\) decreases inspite of strong coupling. Therefore, if we include 
the strong on-site repulsion by an appropriate method and renormalize 
\(E_{\rm F}\) of the quasiparticle, we might obtain a more realistic 
picture of the underdoped cuprates. Finally, the recent experiment has 
clarified that the reduced gap \(2\Delta_0/k_{\rm B}T^*\) is nearly constant 
for the underdoped \({\rm Bi_2Sr_2CaCu_2O_{8+\delta}}\), where \(2\Delta_0\) 
is the magnitude of the superconducting gap estimated by STS and \(T^*\) is 
the temperature at which the in-plane resistivity starts to deviate from the 
\(T\)-linear behavior.~\cite{rf:15} This result suggests the close 
relationship between \(T^*\) and preformed incoherent pairs. 
Consequently, our future problem is to obtain \(2\Delta_0\) consistently and 
study the correlation between \(2\Delta_0\) and 
\(T^*(\simeq T_{\rm c}^{\rm MF})\) or 
\(T_{\rm c}(\simeq T_{\rm c}^{\rm NSR})\) by the use of more realistic 
interaction among \(d\)-electrons in order to explain such experimental 
results.

The authors are grateful to Prof. M. Ido, Prof. M. Oda, Prof. T. Takahashi 
and Dr. T. Yokoya for sharing recent experimental results and fruitful 
discussions. One of the authors (S. K.) is grateful to Dr. T. Hotta for 
his useful comments and discussions. They also thank the Supercomputer 
Center, Kyoto University and the Supercomputer Center, Institute for Solid 
State Physics, University of Tokyo, for the facilities and the use of the 
FACOM VPP500. This work has been supported by a Grant-in-Aid for Scientific 
Research on Priority Areas ``Anomalous Metallic State near the Mott 
Transition'' from the Ministry of Education, Science, Sports and Culture, 
Japan.

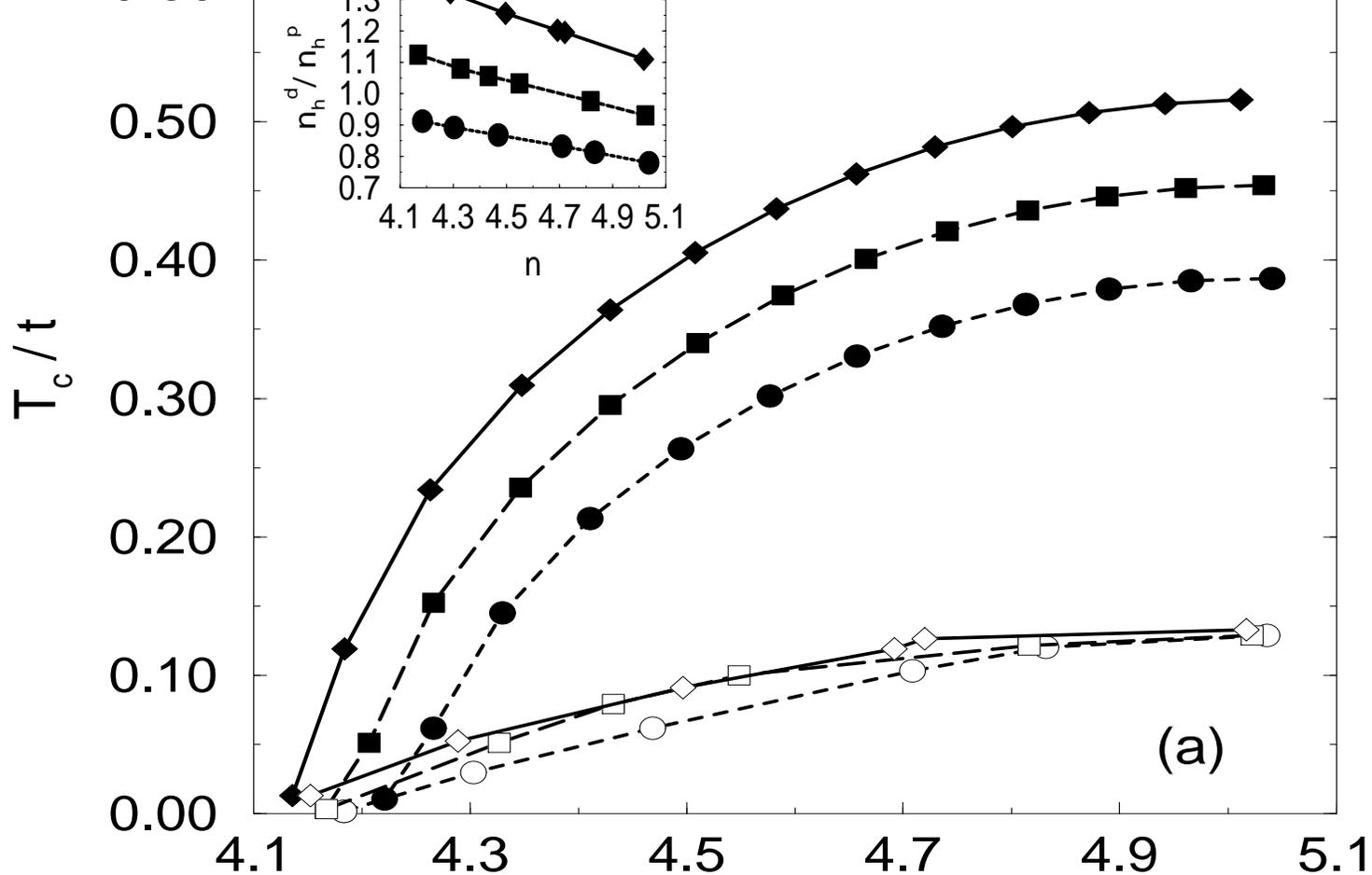
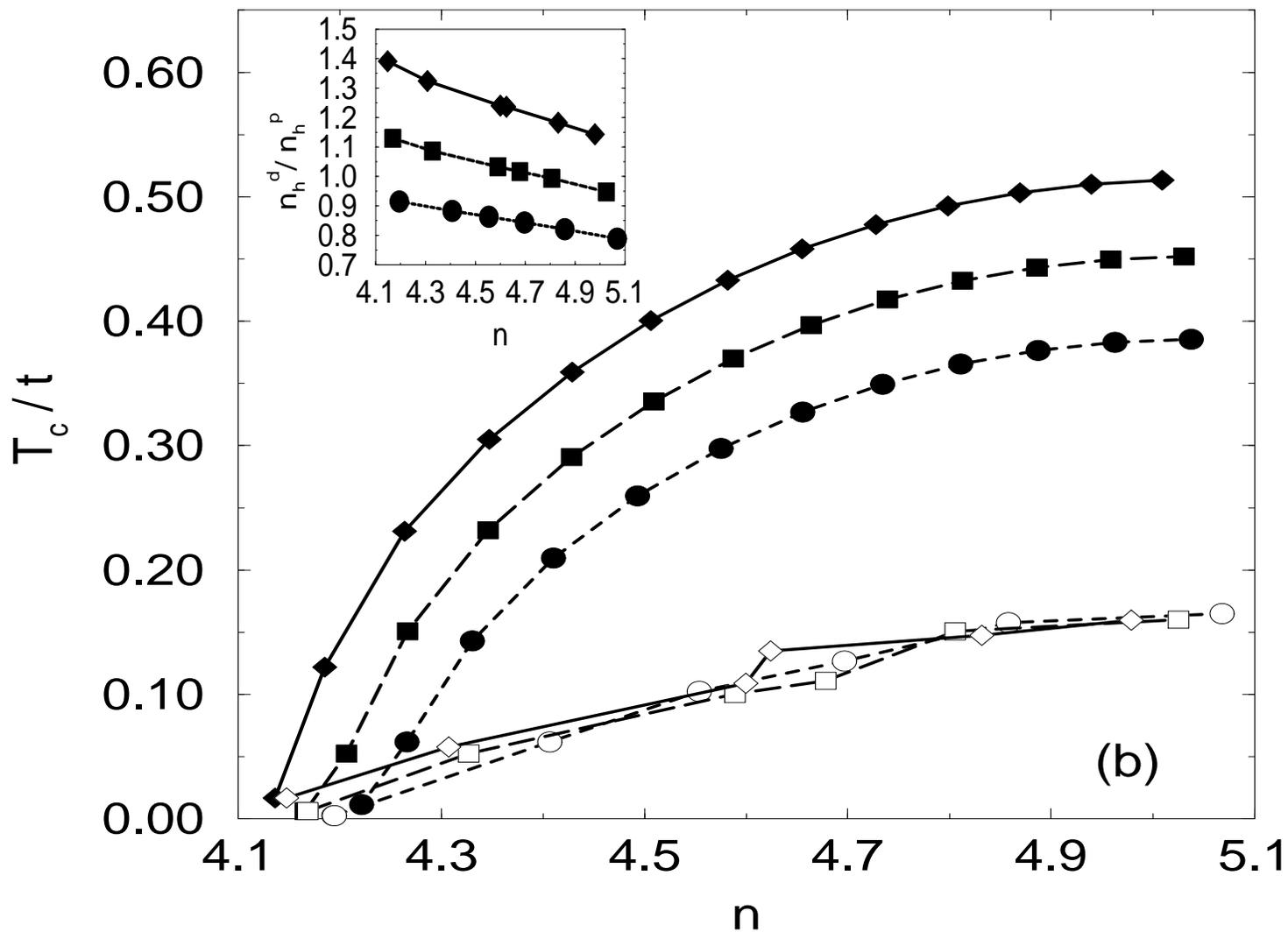
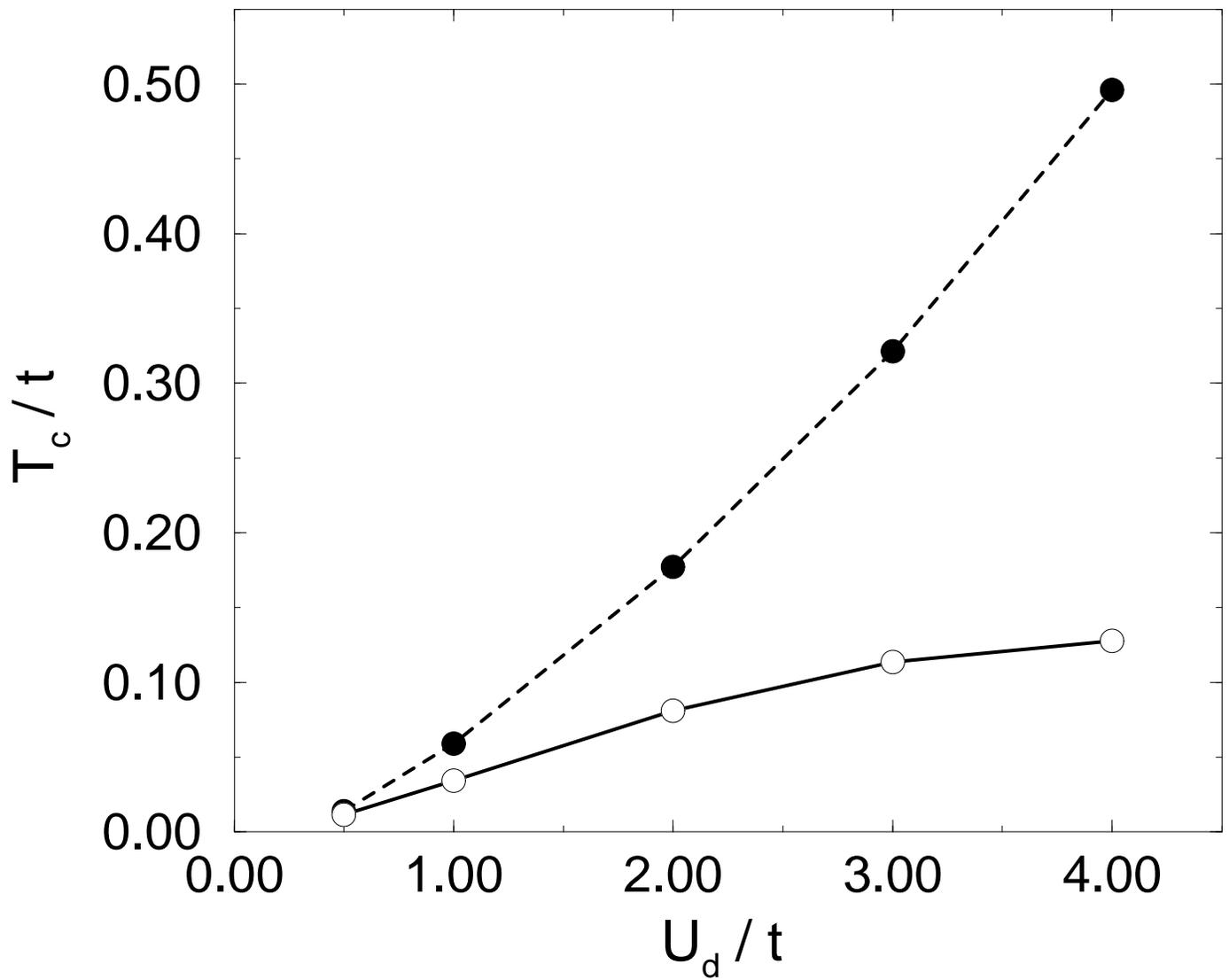
\end{document}